\documentclass[conference]{IEEEtran}
\IEEEoverridecommandlockouts
\usepackage{cite}
\usepackage{amsmath,amssymb,amsfonts}
\usepackage{algorithmic,multirow,enumitem}
\usepackage{graphicx}
\usepackage{textcomp}
\usepackage{xcolor}
\usepackage{algorithm2e}

\def\BibTeX{{\rm B\kern-.05em{\sc i\kern-.025em b}\kern-.08em
    T\kern-.1667em\lower.7ex\hbox{E}\kern-.125emX}}
\begin{document}

\title{Efficiently Processing Workflow Provenance Queries on SPARK}

\author{%
{Rajmohan C.{\small$~^{\#}$}, Pranay Lohia{\small$~^{\#}$}, Himanshu Gupta{\small$~^{\#}$}, Siddhartha Brahma{\small$~^{*}$}, Mauricio Hernandez{\small $~^{\$}$}, Sameep Mehta{\small $~^{\#}$} }%
\vspace{1.6mm}\\
\fontsize{10}{10}\selectfont\itshape
$^{\#}$\,IBM Research - India, $^{*}$\,IBM Research - Almaden, USA, $^{\$}$IBM Watson - USA\\
\fontsize{9}{9}\selectfont\ttfamily\upshape
%
\,\{rajmohanc1, plohia07, higupta8, sameepmehta\}@in.ibm.com, \{brahma, mahernan\}@us.ibm.com \\



}

\maketitle
\vspace{-5pt}

\begin{abstract}
In this paper, we investigate how we can leverage Spark platform for efficiently processing provenance queries on large volumes of workflow provenance data. Existing recursive querying based Spark solutions involve large data scanning cost and hence do not work well for large scale provenance data.  We propose a novel provenance framework which is engineered to quickly determine a small volume of data containing the entire lineage of the queried data-item. This small volume of data is then processed to figure out the provenance of the queried data-item. We study the effectiveness of the proposed framework through experiments on a provenance trace obtained from a real-life unstructured text curation workflow. On provenance graphs containing upto 500M nodes and edges, we show that the proposed framework answers provenance queries in real-time and easily outperforms the naive approaches.
\end{abstract}

\section{Introduction}
\label{sec:ref}

Many applications are encoded as a workflow which executes a sequence of data manipulation operations on raw input data. Provenance is an important requirement for workflow management systems as it enables various use-cases e.g., data-quality, compliance, problem diagnosis etc. For example, if the value of a data-item is erroneous, we can examine its lineage to investigate which transformation has introduced the error and hence fix this transformation. In this paper, we present efficient Spark algorithms for processing large scale workflow provenance data and answer lineage queries.

For a representative example, consider the table Person1. Numbers in bracket represent an id assigned to each attribute-value. Next consider a transformation R1 which filters out persons with age less than 25 and populates the table Person2. Values for attributes \textit{Name}, \textit{City} and \textit{Age} in tuples T5, T6 and T7 are hence derived from values for attributes \textit{Name}, \textit{City} and \textit{Age} in tuples T1, T2 and T3 respectively. Further consider a transformation R2 which works on table Person2 and computes the average age of persons in each city. The resulting output is shown in Table~\ref{tab:p3}. The value for attribute \textit{City} in tuple T8 is derived from values of attribute \textit{City} in tuples T5 and T6. Similarly the value for attribute \textit{Age} in tuple T8 is derived from values of attribute \textit{Age} in tuples T5 and T6. Values for attributes \textit{City} and \textit{Age} in tuple T9 is derived from one value each - value of attribute \textit{City} and \textit{Age} in tuple T7. The workflow provenance data captures these lineages among input and output attribute-values across each transformation, as they are executed. . 

\hspace{-11pt}\textbf{Provenance Data Model:} We assume that the provenance data is specified as a set of triples $\langle$$src$, $dst$, $op$$\rangle$  where $src$ and $dst$ represent the ids of the parent and child data-items and $op$ represents the transformation applied along with any metadata (e.g., run-time parameters, timestamp etc). Table ~\ref{tab:p7} shows the provenance data associated with the representative example.  We also visualize the provenance data as a directed acyclic graph $G(V,E)$ wherein data-items (i.e., $src$ and $dst$) in provenance triples form the vertices $V$ and the provenance triples form the edges $E$ (Table~\ref{tab:p8}).

\begin{table}[!tb]
    \begin{minipage}{.34\linewidth}
      \caption{Person1}
      \vspace{-9pt}
     \label{tab:p0}
      \centering
      {
 \renewcommand{\tabcolsep}{3pt}
      \scalebox{0.7}{
        \begin{tabular}{cccc}
  
\hline
& Name & City &  Age \\
            \hline
\scriptsize{T1} & Steve (1) & NY (2) & 30 (3) \\
\hline
{\scriptsize T2} & Mark (4) & NY (5) &  40 (6)  \\
\hline
T3 & Shane (7) & LA (8) &  40 (9) \\
\hline
T4 & Mary (10)& NY (11)&  20 (12) \\
\hline
&&&\\
        \end{tabular}}
    }    
    \end{minipage}%
    \begin{minipage}{.34\linewidth}
      \caption{Person2}
      \vspace{-9pt}
      \label{tab:p1}
      \centering
      {
     \renewcommand{\tabcolsep}{3pt}
       \scalebox{0.7}{
        \begin{tabular}{cccc}
\hline 
& Name & City & Age \\
            \hline
T5 & Steve (13) & NY (14) & 30 (15) \\
\hline
T6 & Mark (16) & NY (17) &  40 (18) \\
\hline
T7 & Shane (19) & LA (20) &  40 (21)  \\
\hline
&&\\
&&\\
        \end{tabular}}}
    \end{minipage} 
 \begin{minipage}{.22\linewidth}
      \caption{AvgAge}
        \vspace{-9pt}
      \label{tab:p3}
      \centering
      {\scriptsize
     \renewcommand{\tabcolsep}{3pt}
	\scalebox{0.7}{
        \begin{tabular}{ccc}
            \hline
& City & Age \\
\hline
T8 & NY (22) & 35 (23) \\
\hline
T9 & LA (24) & 40 (25) \\
\hline
        \end{tabular}}}
    \end{minipage}%
\vspace{-20pt} 
\end{table}

\vspace{-10pt}

\begin{table}[!ht]
\vspace{-5pt}
\begin{minipage}{.4\linewidth}
\setlength\tabcolsep{0.8pt}
\caption{Provenance Data}
\vspace{-9pt}
\label{tab:p7}
\centering
 \renewcommand{\tabcolsep}{5pt}
\scalebox{0.7}{
\begin{tabular}{ccccc}
\hline
src & dst & op & ccid\\
\hline
1 & 13  & R1 & 1\\
4 & 16  & R1 & 2\\
7 & 19  & R1 & 3\\
2 & 14  & R1 & 4\\
5 & 17  & R1 & 4\\
14 & 22  & R2 & 4\\
17 & 22 &  R2 & 4\\
8 & 20 &  R1 & 5\\
20 & 24 & R2 & 5\\
3 & 15  & R1 & 6\\
6 & 18  & R1 & 6\\
15 & 23 &  R2 & 6\\
18 & 23 &  R2 & 6\\
9 & 21 & R1 & 7\\
21 & 25 & R2 & 7\\
\hline
&&&\\
        \end{tabular}}
\end{minipage}
\begin{minipage}{.55\linewidth}
\setlength\tabcolsep{0.8pt}
\caption{Provenance Graph}
\vspace{-10pt}
\label{tab:p8}
\scalebox{0.8}{
\begin{tabular}{c}
\includegraphics[width=4.5cm]{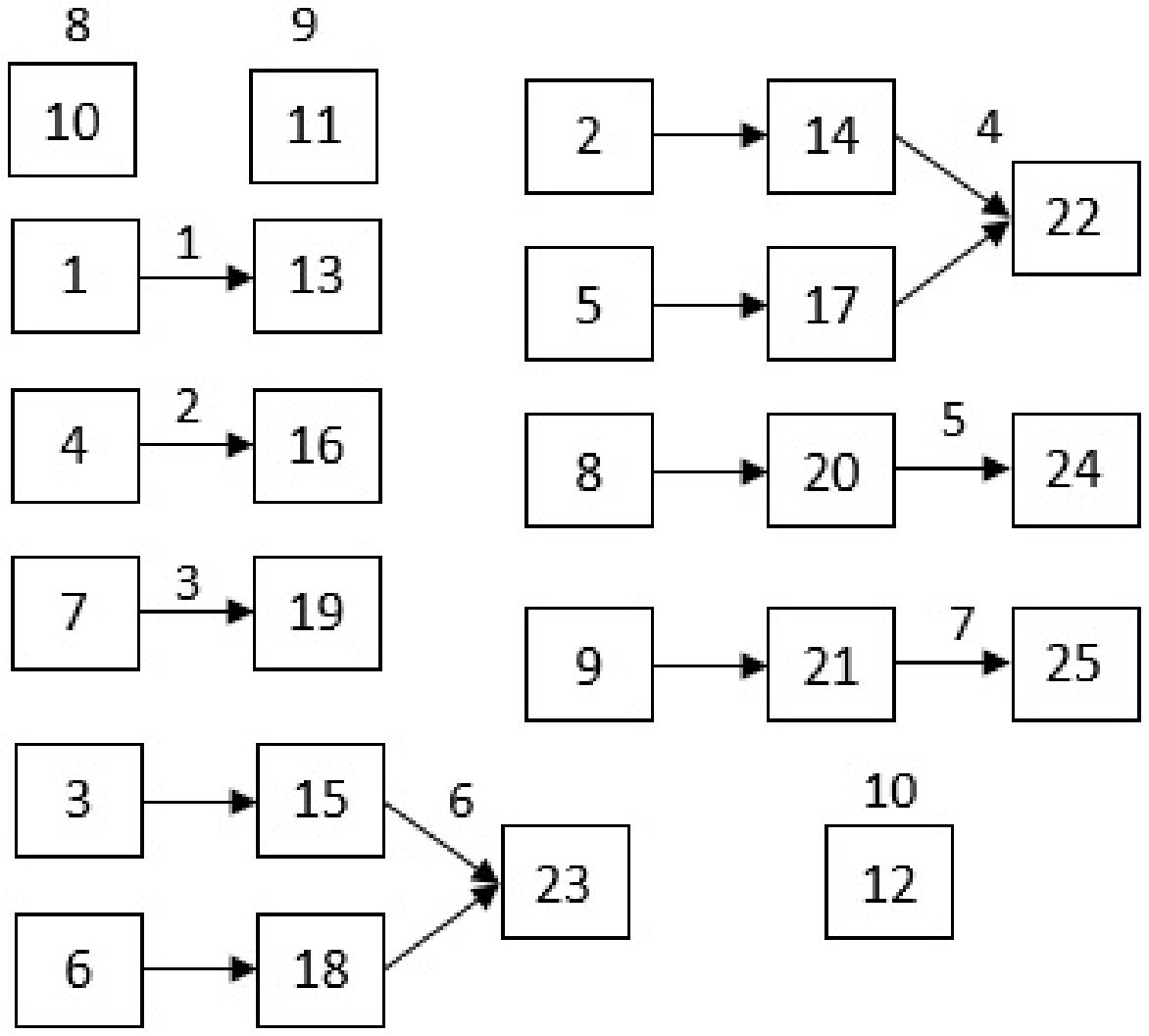}
        \end{tabular}}
\end{minipage}

\vspace{-15pt}
\end{table}

\hspace{-13pt} \textbf{Provenance Query:}  Given a query data-item $q$, we want to track its lineage i.e., all its ancestors and the details of all transformations involved. For example, lineage of data-item 23 (i.e., the value of attribute $Age$ of tuple $T8$ in entity $AvgAge$) will return that data-item 23 is derived from data-items 15 and 18 via transformation R2 and data-items 15 and 18 are derived from data-items 3 and 6 respectively via transformation R1. 
\\
\textbf{Contributions:} A naive approach to answer a provenance query is to recursively process the provenance data. We start with the queried data-item $q$, find those provenance triples which describe its immediate lineage and obtain its parents. We then find the parents of $q$'s parents  and follow this process until we can no longer trace the lineage further.  This approach is adopted by many systems e.g., Trio~\cite{Trio}, GridDB~\cite{Liu-vldb04}, Titian~\cite{Titian} etc. This obviously takes time as we need to issue many queries.  Secondly, as Spark does not support indexing, Spark needs to scan the data to find the parents of a data-item.  This hence does not scale for large volumes of data. A second approach is to pre-compute and materialize the transitive closure of the lineage dependencies (i.e., the provenance of each data-item). This allows retrieval of a data-item's lineage using a single query. However this results in a huge increase in the storage cost as the information regarding common ancestors gets replicated multiple times.  This approach hence also does not scale.

In this paper, we propose a novel approach wherein we first quickly determine a small volume of data which contains the entire provenance output of the queried data-item. We then extract and recursively query this small volume of data. As the recursive querying happens on a small volume of data, we do not incur a large data processing cost. Contributions of this paper are hence as follows.

\begin{itemize}
\item We propose a novel provenance framework wherein we first compute weakly connected components in provenance graph and further partition the large components as a collection of weakly connected sets (section 3). We then effectively navigate the weakly connected components and sets, thus computed, to determine a minimal volume of data containing the entire provenance output of the queried data-item (section 3 and 4).
\item We propose a novel provenance graph partitioning approach wherein we exploit the workflow dependency graph to recursively partition the large components in workflow provenance graph (section 3). 
\item Our experiments on provenance graphs obtained from a real-life text curation workflow and containing upto 500M nodes and edges show that the proposed approaches significantly beat the naive approaches (section 5). The performance is realtime if all data can be cached in RAM.
\item Space overheads are (1) storing two set-ids with each provenance triple and (2) storing set-dependencies i.e., how the sets derive each other. The number of set dependencies are upper-bounded by the number of provenance triples and in practice, are only a small fraction of it. The proposed framework hence has a minimal space overhead.
\end{itemize}

\vspace{-10pt}
\section{Background}

\hspace{-10pt}\textbf{Apache Spark :} Spark uses the resilient distributed data set (RDD) as its basic data type.  An RDD partitions the data across the cluster nodes. In this paper, we will be mainly concerned with Spark \textit{filter} and \textit{lookup} operations. The \textit{filter} operation scans each row of an RDD and checks whether the filter conditions are satisfied or not. A \textit{lookup} is a specific kind of filter where one or more columns are checked for equality. To accelerate \textit{lookup} operations, we can hash-partition an RDD on one or more columns and, this process moves all rows with same key to one partition.  With hash-partitioning enabled, a lookup needs to scan only one partition. Hash-partitioning also accelerates \textit{filter} performance, if the filter conditions involve checking column equality on hashed columns. RDDs can also be cached and this avoids re-computation of an RDD, each time it is accessed.
\\
\textbf{Weakly Connected Sets and Components:} A semipath joining vertices $u_1$ and $u_k$ in a directed graph $G$=($V$,$E$) is a sequence of vertices $u_1$,$u_2$$\ldots$,$u_k$ s.t. for each $i$, 1$\leq$$i$$\leq$$k-1$ either there exists an edge $u_i \rightarrow u_{i+1}$ in $E$ or there exists an edge $u_{i+1} \rightarrow u_i$ in $E$.  A set of vertices $W$$\subseteq$$V$ is called weakly connected if there exists a semipath between each vertex pair in $W$. A maximal weakly connected set of vertices is a weakly connected component in $G$. 
\\
\textbf{Notation:} We use the terms ``connected component" and ``connected set" as a shorthand for ``weakly connected components" and ``weakly connected sets", though they are different abstractions in graph theory.  
\vspace{-5pt}
\section{The Provenance Framework}
\label{sec:ccpf}
\vspace{-3pt}
\subsection{Recursive Querying on Spark (RQ)}
\label{sec:rqs}
We first discuss the challenges in executing recursive querying (RQ) on Spark. Let us denote the provenance data RDD as \textit{provRDD}. As discussed, RQ involves executing many queries to trace the entire lineage of a data-item $q$. The number of queries are equal to the length of the largest provenance path in the lineage of data-item $q$. Each such query involves finding parents of one or more data-items $\mathcal{I}$.  As discussed above, if we hash-partition the \textit{provRDD} on field \textit{dst}, this moves all provenance triples with the same \textit{dst} field to one partition and we can hence find the parents of a data-item by scanning one partition of \textit{provRDD}. To find parents of all data-items in $\mathcal{I}$, we need to scan at most $|\mathcal{I}|$ number of partitions. This is because, some data-items in $|\mathcal{I}|$ may be in the same partition and the parents of these data-items can hence be obtained by scanning this partition only once.  If the lineage size (i.e., number of ancestors) of queried data-item $q$ is $N$, we hence require scanning a maximum of $|N|$ number of partitions. The overall RQ cost will hence depend upon the number of queries executed, set of lookups made as part of each query,  and the distribution of field \textit{dst} across the \textit{provRDD}.

\vspace{-2pt}
\subsection{Connected Components and Provenance} 
\label{sec:ccprov}
\vspace{-2pt}

We observe that the workflow provenance graph, formed by attribute-values, is a large collection of weakly connected components. This is because many attribute-values do not share any common ancestors. This is best evidenced by looking at Table~\ref{tab:p8} which shows the provenance graph for the representative example. This graph contains 10 weakly connected components. We notice that a data-item and all its ancestors as well as descendants, share the same weakly connected component. This property can be used to speed up the processing of provenance queries. Given a queried data-item $q$, we first find out its weakly connected component id and then retrieve all provenance triples in this component.  We then process the triples in this component recursively to figure out the provenance of data-item $q$. As the size of a component is much smaller than the whole provenance graph, the recursive querying executes faster. We hence compute weakly connected components on the provenance graph and then append the connected component id with each provenance triple as shown in Table~\ref{tab:p7}. This computation is part of pre-processing and needs to be done only once. 

{
\begin{algorithm}[t]
 \label{alg:ccprov}
 \footnotesize{\KwIn{Hash-Partitioned Provenance RDD \textit{provRDD}, data-item $q$}
 \KwOut{Lineage of data-item $q$}
 $c$ $\leftarrow$ Find-Connected-Component(\textit{provRDD}, $q$)\;
 $c\_$\textit{provRDD} $\leftarrow$ Find-ProvTriples-In-Component(\textit{provRDD}, $c$)\;
 return Recursive-Query($c\_$\textit{provRDD}, $q$)\;
}
 \caption{CCProv}
 \vspace{2pt}
\end{algorithm}
}

Algorithm 1 outlines the algorithm for computing the lineage of a data-item $q$ and it takes the provenance data RDD \textit{provRDD},  hash-partitioned on column $dst$ as input. We first find out the id of the connected component, the data-item $q$ lies in and let it be $c$.  This can be found by scanning a single partition of \textit{provRDD}. We then find all provenance triples in component $c$ and let it be  $c\_$\textit{provRDD}. This is done via a Spark \textit{filter} operation on \textit{provRDD} and this preserves the hash-partitioning logic.  We then recursively process $c\_$\textit{provRDD} to find the lineage of data-item $q$.

\vspace{-4pt}
\subsection{Connected Sets and Provenance} 
\label{sec:csprov}

Though CCProv provides better performance vis-a-vis RQ, it may not be good-enough when the component size is large as CCProv processes large volume of data (i.e., RDD $c\_$\textit{provRDD}). We next discuss CSProv which improves on this aspect. The idea is to pre-process and partition the large components into a collection of weakly connected sets. At query time, we exploit the information regarding how these sets derive each other to quickly find a minimal volume of data containing the entire lineage of the queried data-item. We explain the intuition via a representative example. 

\begin{table}[!t]
   \vspace{-15pt}
    \begin{minipage}{.5\linewidth}
      \caption{A Component C}
     \label{tab:acc}
     \vspace{-9pt}
     \centering
      \begin{tabular}{c}
\includegraphics[width=4.5cm]{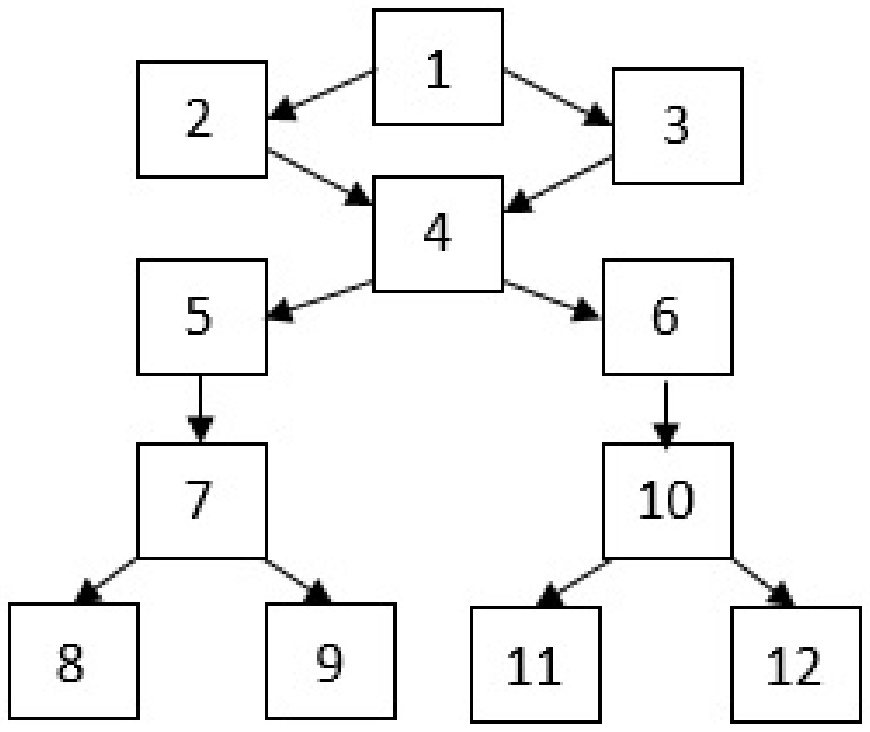}
        \end{tabular}
    \end{minipage}%
    \begin{minipage}{.5\linewidth}
      \caption{Provenance Data}
      \label{tab:acc_new_edge}
      \vspace{-9pt}
      \centering
    {\scriptsize
   \scalebox{0.7}{
   \begin{tabular}{ccccc}
\hline
src & dst & op & src\_csid & dst\_csid\\
\hline
1 & 2 &  - & S1 & S1\\
1 & 3 &  - & S1 & S1\\
2 & 4 &  - & S1 & S2\\
3 & 4 &  - & S1 & S2\\
4 & 5 &  - & S2 & S2\\
4 & 6 &  - & S2 & S2\\
5 & 7 &  - & S2 & S3\\
7 & 8 &  - & S3 & S3\\
7 & 9 &  - & S3 & S3\\
6 & 10 &  - & S2 & S4\\
10 & 11 &  - & S4 & S4\\
10 & 12 &  - & S4 & S4\\
\hline
&&&\\
        \end{tabular}}}

      \caption{Set Dependencies}
      \label{tab:acc_ccedges}
      \vspace{-9pt}
\centering
 {\scriptsize
 \scalebox{0.7}{
\begin{tabular}{cc}
\hline
 src\_csid & dst\_csid\\
\hline
S1 & S2 \\
S2 & S3 \\
S2 & S4 \\
\hline
\end{tabular}}}
    \end{minipage}%
 \vspace{-8pt}
\end{table}

{
\begin{algorithm}[th]
 \label{alg:csprov}
 \footnotesize{
 \KwIn{Hash-Partitioned Provenance RDD \textit{provRDD}, Hash-Partitioned Set Dependencies RDD \textit{setDepRDD}, data-item $q$}
 \KwOut{Lineage of data-item $q$}
 $cs$ $\leftarrow$ Find-Connected-Set(\textit{provRDD}, $q$)\;
 $\mathcal{S}$ $\leftarrow$ $cs$ $\cup$ Find-Set-Lineage(\textit{setDepRDD}, $cs$)\;
 \textit{cs\_provRDD} $\leftarrow$ $\phi$\;
 \For{connected set $s$ in $\mathcal{S}$}{
	 \textit{cs\_provRDD} $\leftarrow$  \textit{cs\_provRDD}  $\cup$ Find-ProvTriples-With-DerivedItem-In-Set(\textit{provRDD},$s$)\;
 }
return Recursive-Query(\textit{cs\_provRDD}, $q$)\;
}
 \caption{CSProv}
 \vspace{2pt}
\end{algorithm}
}

Consider a weakly connected component C as shown in Table~\ref{tab:acc}. Consider, we partition the component C in 4 weakly connected sets  - S1, S2, S3 and S4. These sets are formed by data-items \{1, 2, 3\},  \{4, 5, 6\},  \{7, 8, 9\} and  \{10, 11, 12\} respectively. We also maintain the set dependencies - how these sets contribute to the derivation of other sets. The set S1 contributes to the derivation of set S2 as data-items 2 and 3 in set S1 derive data-item 4 in set S2. Set S2 derives set S3 as data-item 5 in set S2 derives data-item 7 in set S3. Similarly set S2 derives set S4. Note that sets S3 and S4 do not contribute to the derivation of any set (Table~\ref{tab:acc_ccedges}).

Consider that we query the provenance of data-item 8. This belongs to the set S3. From set-dependencies, we find that set S2 derives set S3 and set S1 derives set S2. Hence sets S1 and S2 are relevant to the derivation of set S3. These three sets together contain all ancestors of the data-item 8. We only process those triples whose derived ($dst$) data-item is in sets S1, S2 and S3.  We do not need to process set S4 triples as the set-dependencies tell us that set S4 neither contributes to the derivation of set S3 nor to the derivation of any ancestor set of set S3. We hence end up processing a smaller volume of data, in this example 3 less provenance triples.  

CSProv requires the following updates on the provenance data model discussed in section~\ref{sec:ref}.
\begin{itemize}
\item \textbf{Provenance Data:} Data-items \textit{src} and \textit{dst} in a provenance triple may lie in two different weakly connected sets and we hence maintain the set id of both items. We add the columns \textit{src\_csid} and \textit{dst\_csid} in the schema and drop the field \textit{ccid} from the provenance triple (Table~\ref{tab:acc_new_edge}).
\item \textbf{Set Dependencies:} We also maintain how the weakly connected sets are derived from each other (Table ~\ref{tab:acc_ccedges}). We say a set $s_1$ is derived from $s_2$ if there exists at least one data-item $u$ in $s_1$ and at least one data-item $v$ in $s_2$ s.t. there is a provenance triple where $src$ equals $v$ and $dst$ equals $u$.  There are two columns in the schema - \textit{src\_csid} and \textit{dst\_csid} which denote the set-ids of parent and child connected sets. 
\end{itemize}

Algorithm 2 outlines the algorithm CSProv.  It takes provenance data \textit{provRDD} and set dependencies \textit{setDepRDD} as input, both hash-partitioned on field $dst\_csid$. Given queried data-item $q$, we first find out its connected set $cs$. We then construct set $\mathcal{S}$ which includes set $cs$ and its set-lineage i.e., all sets which contribute to the derivation of set $cs$, directly or indirectly. This is done by executing RQ logic on \textit{setDepRDD}. RQ on \textit{setDepRDD} is lightweight due to two reasons. First, the size of \textit{setDepRDD} is likely to be much smaller vis-a-vis \textit{provRDD}. Secondly, the size of set-lineage of set $cs$ is likely to be much smaller than the size of lineage of data-item $q$ and hence much smaller number of queries need to be executed.   

For each set $s$ in $\mathcal{S}$ , we find the provenance triples s.t., the data-item $dst$ is in connected-set $s$.  As \textit{provRDD} is hash-partitioned on field $dst\_csid$, this requires scanning at most $|\mathcal{S}|$ number of partitions. As discussed, the size of set $\mathcal{S}$ is small and this operation is hence light-weight as well. A union of all these provenance triples i.e., $cs\_$\textit{provRDD} contains the entire lineage of data-item $q$. We then recursively process $cs\_$\textit{provRDD} to compute the lineage of data-item $q$. Again, the size of $cs\_$\textit{provRDD} is likely to be much smaller that the size of the component, the queried data-item lies in. Recursive querying on $cs\_$\textit{provRDD} is hence light-weight as well. 

Note that when the queried data-item $q$ lies in a small component $c$, CSProv reduces to CCProv. Small components are not partitioned and each small component is managed as a single weakly connected set (i.e., itself). The set $\mathcal{S}$ hence only contains the set/component $c$. 
\section{Partitioning Large Components}
\label{sec:partition}

\begin{figure*}[t]
\includegraphics[width=\linewidth, height=3.5cm]{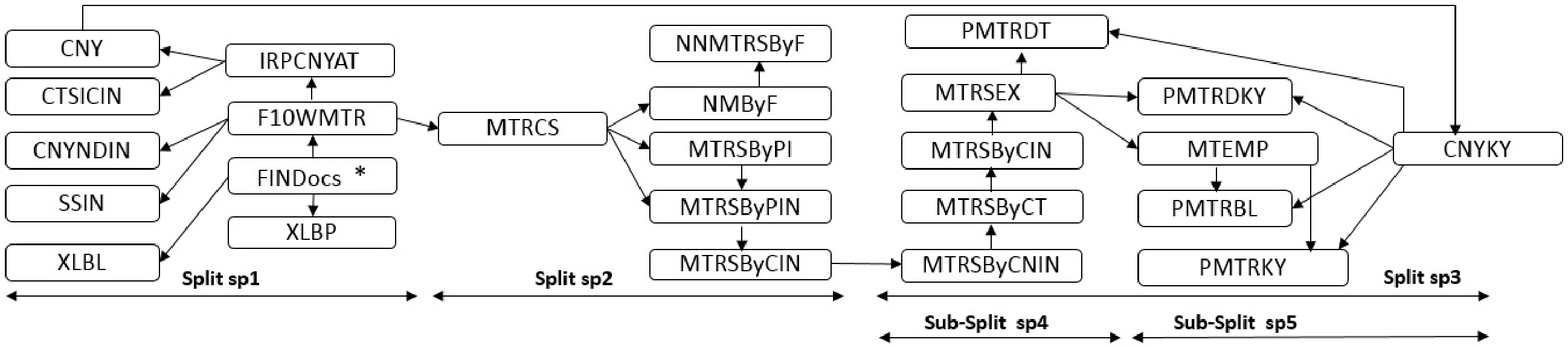}
 \vspace{-15pt}
  \caption{The Text Curation Workflow}
  \vspace{-10pt}
  \label{fig:wcs}
\vspace{-10pt}
\end{figure*}

In section~\ref{sec:csprov}, we identified the following criteria for algorithm CSProv to work efficiently. 
\begin{itemize}
\item C1 - Number of set-dependencies should be small.
\item C2 - The set-lineage of a set should be small.
\item C3 - The size of each connected set should be small.
\end{itemize}
Criteria C1 and C2 imply that CSProv can construct the set-lineage of a set $cs$ cheaply. Criteria C2 and C3 imply that small number of triples (i.e., the size of $cs\_$\textit{provRDD}) need to be recursively processed. We next discuss how we partition the large components, so as the resulting sets satisfy these criteria. We exploit the workflow dependency graph for the same. The dependency graph specifies dependencies among the tables and hence an order in which various tables are generated e.g., the dependency graph in Figure~\ref{fig:wcs} specifies that the table \textit{MTRCS} can be generated only after table \textit{F10WMTR} is generated. We first develop the following notation. 
\\
\textbf{Notation:} Let $G_{wf}$ represent the workflow dependency graph. Let a split be a sub-set of tables in dependency graph $G_{wf}$ s.t., these tables are weakly-connected in graph $G_{wf}$. Figure~\ref{fig:wcs} shows a partitioning of the dependency graph across three splits - $sp1$, $sp2$, $sp3$. Note that the tables in each split are weakly-connected. Let $V(sp,c)$ be the set of those vertices in provenance graph $G(V,E)$ which belong to component $c$ and belong to a table in split \textit{sp}. Let $G[V(sp,c)]$ be the subgraph induced by the vertices $V(sp,c)$. We also call $G[V(sp,c)]$ the provenance subgraph induced by split $sp$ and component $c$. Let $W(sp,c)$ be the set of weakly connected components in subgraph $G[V(sp,c)]$.

{\small
\begin{algorithm}[t]
 \label{alg:plc}
 \footnotesize{
 \KwIn{Large Component $c$, Set of weakly-connected-splits $S$}
 \KwOut{Set of Weakly-Connected-Sets}
 $W$ $\leftarrow$ $\phi$\;
 \For{split $sp$ in $S$}{
   $W(sp,c)$ $\leftarrow$ Compute-Weakly-Connected-Components($G[V(sp,c)]$)\;
   \For{component $cn$ in $W(sp,c)$}{
       \eIf{node-count of $cn$ $\geq$  $\theta$}{
           $SS$ $\leftarrow$ Get-Weakly-Connected-Sub-Splits($sp$)\;
           $W$ $\leftarrow$ $W$ $\cup$ Partition-Large-Component($cn$, $SS$)\;
       }{
           $W$ $\leftarrow$ $W$ $\cup$ cn\;
       }
   }
 }
 return $W$\;
 \vspace{3pt}}
 \caption{Partition-Large-Component}
 \vspace{-5pt}
\end{algorithm}
}
\vspace{5pt}
Algorithm 3 outlines the details. We first partition the dependency graph $G_{wf}$ into a set of disjoint splits $S$. Algorithm \textit{Partition-Large-Component} takes a large component $c$ and the dependency graph splits $S$ as input, and returns the set of weakly connected sets $W$ as output. For each split $sp$ in $S$, we first construct the subgraph $G[V(sp,c)]$ and then compute the weakly connected components $W(sp,c)$ in it. The procedure then iterates over each component $cn$ in $W(sp,c)$. If the number of vertices in component $cn$ is less than a threshold $\theta$, it is not processed further and is inserted in the output set $W$. If not, we further partition split $sp$ into a set of disjoint and weakly connected sub-splits $SS$ and recursively call the procedure \textit{Partition-Large-Component} with component $cn$ and split-set $SS$ as input. 
\\
\textbf{Computing Set Dependencies:} After all large components are partitioned, the fields \textit{src\_csid} and \textit{dst\_csid} associated with each provenance triple are populated using the connected sets, thus generated. We then find those provenance triples wherein the columns $src\_csid$ and $dst\_csid$ take different values. The set of distinct ($src\_csid$, $dst\_csid$) pairs in such triples, form the set dependencies.
\\
\\
\textbf{Discussion:} The constraint that all tables in each split are weakly connected, is a key part of the algorithm.  Note that for any given large component $c$ and split $sp$, no two components in $W(sp,c)$ contribute a set-dependency i.e., there is no set-dependency ($cn_1$, $cn_2$) s.t. both $cn_1$ and $cn_2$ are in $W(sp,c)$. This is because, the set $W(sp,c)$ is obtained by computing weakly connected components on subgraph $G[V(sp,c)]$ and any two components in $W(sp,c)$ are hence, by definition, disconnected. This ensures that the number of set-dependencies are small (criteria C1). Secondly, this increases the likelihood that a data-item's local lineage (i.e., its few immediate ancestors) can be found in the same weakly connected set, this data-item lies in and hence only few sets returned by the procedure  are relevant to the lineage of a queried data-item (criteria C2). Finally the condition that the size of each set has to be less than a threshold $\theta$, ensures that the size of each set is small (criteria C3). 

Note that, if we consider each table in dependency graph as a separate split, CSProv reduces to RQ. Each attribute-value is a connected component and  provenance triples capture the set dependencies. If we consider all tables in dependency graph as part of one split, CSProv reduces to CCProv.

\section{Experimental Evaluation}
\label{sec:exp}

\hspace{-10pt}\textbf{Provenance Data Set:} We used a provenance trace obtained from a real-life workflow deployed in our lab for creating financial domain knowledge-bases~\cite{Bharadwaj17}.  The workflow parses SEC filing documents~\cite{sec}. Each SEC document contains data pertaining to many thousands of financial metrics and the workflow curates this data.  Figure~\ref{fig:wcs} shows the workflow dependency graph comprising 25 entities (tables). For each entity, we have only shown its acronym so as to remove any confidential information. The workflow contains various transformations involving entity annotation, extraction and resolution. For each transformation, the lineage relationships among the child and parent attribute-values are captured. The workflow contains many UDFs and the lineage service assumes that each attribute-value in an UDF output is dependent on each attribute-value in the UDF input. The entity \textit{FINDocs} (marked *) forms the workflow input.

This workflow is executed on a set of 532 financial documents. The obtained provenance trace is 1.6GB in size and contains 6.4M triples with 4.6M attribute-values.  The provenance graph hence contains 4.6M nodes and 6.4M edges. These attribute-values have widely different derivation patterns. 32 attribute-values are being directly derived from more than 100 parent values, with the maximum being 450. 3963 values are directly derived from more than 10 parents but less than 100 parents. Rest of the attribute-values have less than 10 parents.  
\\
\textbf{Spark Cluster} The cluster runs Spark v2.0.2, has 8 nodes with 12 cores each, 2.4GHz processor and 120 GB RAM.  
\\
\\
\textbf{Weakly Connected Components:} We computed weakly connected components in the provenance graph, using Spark implementation provided at ~\cite{cc-spark-implementation} and it took 6 mins to compute them. Three of these components are large containing 1.2M, 0.9M and 0.7M nodes, and 2.7M, 1.4M and 1.2M edges (triples) respectively.  We denote these three large components by notations LC1, LC2 and LC3 respectively. 132 components contain between 910 and 7453 nodes. Rest of the components have 100 or lesser number of nodes.
\begin{table}[t!]
  \begin{center}
    \caption{Weakly Connected Sets Statistics}
    \label{tab:wcsets}
    \vspace{-8pt}
   {\small
   \scalebox{0.8}{
    \begin{tabular}{|c|c|c|c|c|c|c|c|c|c|}
    \hline
    \multicolumn{10}{|c|}{Number of sets, \# sets with $\geq$ 1000 nodes, \# nodes in largest set}\\
  \hline
     & \multicolumn{3}{|c|}{Split sp1} & \multicolumn{3}{|c|}{Split sp2} & \multicolumn{3}{|c|}{Split sp3} \\
  \hline
     LC1 &  \multicolumn{3}{|c|}{20, 0, 490}   &   \multicolumn{3}{|c|}{29696, 4, 21734} &  \multicolumn{3}{|c|}{219879, 11, 3291} \\
  \hline
  LC3  &  \multicolumn{3}{|c|}{10, 0, 313} & \multicolumn{3}{|c|}{15491, 1, 2578} & \multicolumn{3}{|c|}{128264, 0, 643}\\
\hline
  LC2  &  \multicolumn{3}{|c|}{1, 0, 4}  & \multicolumn{3}{|c|}{1,0,211} & \multicolumn{3}{|c|}{1, 1, 0.9M}\\
\hline
       \multirow{2}{*}{LC2\_lc1} & \multicolumn{6}{|c|}{Split sp4} & \multicolumn{3}{|c|}{Split sp5}  \\
\cline{2-10}
    & \multicolumn{6}{|c|}{64737, 0, 30} & \multicolumn{3}{|c|}{132599, 2, 24733}\\
  \hline
    \end{tabular}}
  }
  \end{center}
\vspace{-18pt}
\end{table}

\begin{table*}
\centering
\begin{minipage}{0.3\linewidth}
\caption{\footnotesize{Class SC-SL Query Times (s)}}
\vspace{-10pt}
\label{tab:scsl}
  {\small
   \scalebox{0.8}{
    \begin{tabular}{|c|c|c|c|c|}
    \hline
         & 10M & 100M & 250M & 500M \\   
    \hline
       RQ & 2.3 & 8.9 & 10.8 & 16.5  \\
    \hline   
       CCProv & 0.3 & 0.4 & 0.6 & 0.9  \\
    \hline   
       CSProv & 0.3 & 0.4 & 0.6 & 0.9 \\
    \hline   
    \end{tabular}
   }}
\end{minipage}
\begin{minipage}{0.3\linewidth}
\caption{\footnotesize{Class LC-SL Query Times (s)}}
\vspace{-10pt}
\label{tab:lcsl}
  {\small
    \scalebox{0.8}{
    \begin{tabular}{|c|c|c|c|c|}
    \hline
         & 10M & 100M & 250M & 500M \\   
    \hline
       RQ & 2.1 & 8.3 & 11.4 & 16.0  \\
    \hline   
       CCProv & 2.3 & 5.0 & 6.2 & 7.9  \\
    \hline   
       CSProv & 0.6 & 0.8 & 1.1 & 1.6 \\
    \hline   
    \end{tabular}}
   }
\end{minipage}
\begin{minipage}{0.3\linewidth}
\caption{\footnotesize{Class LC-LL Query Times (s)}}
\vspace{-10pt}
\label{tab:lcll}
  {\small
    \scalebox{0.8}{
    \begin{tabular}{|c|c|c|c|c|}
    \hline
         & 10M & 100M & 250M & 500M \\   
    \hline
       RQ & 2.7 & 9.1 & 12.7 & 20.0  \\
    \hline   
       CCProv & 2.5 & 5.5 & 7.0 & 9.1  \\
    \hline   
       CSProv & 0.8 & 1.3 & 1.7 & 2.2 \\
    \hline   
    \end{tabular}}
   }
\end{minipage}
\vspace{-10pt}
\end{table*}

\begin{table}
\caption{\footnotesize{RDDs Cached on Disk, Class LC-LL Query Times (s)}}
\vspace{-10pt}
\label{tab:lcll-disk}
\centering
  {\small
    \scalebox{0.8}{
    \begin{tabular}{|c|c|c|c|c|}
    \hline
         & 10M & 100M & 250M & 500M \\   
    \hline
       RQ & 7 & 20 & 47 & 101  \\
    \hline   
       CCProv & 5.5 & 9 & 16 & 31 \\
    \hline 
       CSProv & 3 & 6 & 11 & 17 \\
    \hline
    \end{tabular}}
   }
\vspace{-10pt}
\end{table}

\hspace{-10pt}\textbf{Weakly Connected Sets:} We next partitioned the three large components using Algorithm 3. We partitioned the workflow dependency graph $G_{wf}$ in three weakly connected splits $sp1$, $sp2$, $sp3$ as shown in Figure~\ref{fig:wcs}. We set threshold $\theta$ to 25K nodes. Table~\ref{tab:wcsets} presents the statistics on the connected sets obtained. For each large component $c$ and for each split $sp$, we note - (a) number of sets computed i.e., $|W(sp,c)|$, (b) number of sets in $W(sp,c)$ with  $\geq$ 1000 nodes and (c) number of nodes in the largest set in $W(sp,c)$ (i.e. the set containing maximum nodes).

The component LC1 got partitioned in a total of 249595 weakly connected sets with splits $sp1$, $sp2$ and $sp3$ accounting for 20, 29696 and 219879 sets  respectively. Largest sets in $W(sp1,LC1)$, $W(sp2,LC1)$ and $W(sp3,LC1)$ turned out to contain 490, 21734 and 3291 nodes respectively and hence did not need not further partitioning. The component LC3 got partitioned in 143765 sets with the largest sets in $W(sp1,LC3)$, $W(sp2,LC3)$ and $W(sp3,LC3)$ containing 313, 2578 and 643 nodes. No set in $W(sp1,LC3)$, $W(sp2,LC3)$ and $W(sp3,LC3)$ hence required further partitioning.

However, the sub-graph $G[V(sp3,LC2)]$ yielded only a single connected component of size 0.9M. Let us denote it as LC2\_lc1. This component hence needs to be partitioned further. Split $sp3$ is partitioned in two weakly connected sub-splits $sp4$ and $sp5$ as shown in Figure~\ref{fig:wcs} and the procedure  \textit{Partition-Large-Component} is called on component LC2\_lc1 and split-set \{$sp4$,$sp5$\} as input. This time, LC2\_lc1 got partitioned into 197336 sets with sub-splits $sp4$ and $sp5$ accounting for 64737 and 132599 sets respectively. None of these sets contained more than $\theta$ nodes and hence no further partitioning is needed.  Overall, the three large components LC1, LC2 and LC3 get partitioned into 590698 sets and these sets involve 645303 set-dependencies.  Number of these set-dependencies are hence an order of magnitude smaller than the number of provenance triples and the size on disk is 0.03GB. 
\\
\textbf{Scaled Datasets:} We replicated the provenance trace by a factor of 9, 24 and 48 and this generated three scaled provenance graphs containing 100M, 250M and 500M nodes and edges respectively. The sizes on disk are 15, 35 and 71GB respectively. As the data is replicated, these scaled datasets contain 27, 72 and 144 large components respectively. These large components are partitioned and the statistics regarding the resulting sets mirror the stats given in Table~\ref{tab:wcsets}. Number of set dependencies are hence 9, 24 and 48 times vis-a-vis the base dataset and the size on disk are 0.25, 0.67 and 1.3GB respectively. The computation of the connected components and  connected sets on these three scaled datasets took 16, 28 and 50 mins respectively.  
\\
\textbf{Provenance Queries:} We chose three classes of lineage queries to illustrate the effectiveness of proposed approaches. For each class, we chose 10 data-items and queried their lineage using RQ, CCProv and CSProv, on base as well as scaled datasets. The largest provenance path for all LC-LL queries is 10 while it is 7 for all SC-SL and LC-SL queries. 
\begin{itemize}
\item \textbf{SC-SL}: We chose data-items in a small component containing 7453 nodes and 8122 edges. Number of ancestors as well as transformations in lineage of these data-items are between 100 and 200. These queries hence track lineage of data-items with small lineage size. 
\item \textbf{LC-SL} : We chose data-items in large components LC1, LC2, LC3 s.t., both the number of ancestors and transformations in their lineage are between 100 and 200. These queries track lineage of data-items in large components, but with small lineage size.   
\item \textbf{LC-LL}: We chose data-items in large components s.t., both the number of ancestors and transformations in their lineage are between 5000 and 10000. These queries track lineage of data-items in large components, but with considerably larger lineage size vis-a-vis class LC-SL.   
\end{itemize} 

\hspace{-10pt}\textbf{RDDs Cached in RAM:} We first ran experiments with 80 GB executor memory. For all scaled datasets, all RDDs fit in memory with this configuration. The RDDs were hash-partitioned and cached in RAM. All RDDs were loaded with 96 partitions. We executed the lineage queries and measured the average of the time taken by the 10 queries for each class. Tables~\ref{tab:scsl},~\ref{tab:lcsl} and ~\ref{tab:lcll} present the results. We note that CSProv performance is real-time, degrades gracefully with datasize and significantly better than RQ and CCProv.  

\hspace{-10pt}\textbf{RDDs Cached on Disk:} A cluster may not have enough RAM to cache all RDDs.  We next repeated the experiments but cached the hash-partitioned RDDs on disk. Table~\ref{tab:lcll-disk} presents the results. For lack of space, we show results only for class LC-LL. For all steps,  RQ, CCProv and CSProv read the data from disk. As the datasize increases, the gap between RQ and CSProv widens. 
\\
\\
\textbf{Discussion:} We next explain the details of CSProv using a query for each class. One of the 10 data-items queried for LC-SL class, belongs to a connected set in $W(sp3, LC1)$ and this set $cs$ contains 79 nodes and 102 edges. 13 sets in $W(sp2, LC1)$ derive the set $cs$ and these 13 sets are found to be derived from one set in $W(sp1, LC1)$. Set $cs$ and these 14 sets in its set-lineage hence construct the set $\mathcal{S}$ (Algorithm 2), and these 15 sets are found to contain a total of 1816 nodes and 4177 edges. For all datasets, CSProv hence needs to recursively query only 4177 provenance triples while CCProv needs to query 2.7M triples. This leads to the improved performance of CSProv.

A data-item queried for class LC-LL belongs to a connected set $cs$ in $W(sp3, LC1)$ and it contains 3291 nodes and 4403 edges. 4 sets in $W(sp2, LC1)$ derive set $cs$ and these 4 sets are found to be derived from 20 sets in $W(sp1, LC1)$. These 25 sets contain a total of 44196 nodes and 60169 edges. CSProv hence needs to recursively query only 60169 triples while CCProv needs to process 2.7M triples. For SC-SL class, as a small component is not partitioned, both CCProv and CSProv recursively process 8122 triples.  

\hspace{-10pt}\textbf{GraphX:} It is to be noted that GraphX library supports graph-parallel computation on top of Spark but as CCProv/CSProv do not have any graph-parallel computation,  we use core Spark RDDs and not GraphX, for our implementations.

\vspace{-10pt}
\section{Related Work}
\label{sec:rw}

Titian~\cite{Titian} is the only major prior work to have looked at provenance data management and querying on Spark. However, Titian focuses on efficiently capturing provenance data in a Spark workflow.  Once captured, it uses the recursive querying approach to trace the lineage of a record in an RDD. In comparison, our focus is on leveraging Spark platform for efficiently processing provenance data obtained from a workflow management system and not on capturing provenance data in a Spark workflow. We propose a novel framework for optimizing workflow provenance queries on Spark which exploits the workflow dependency graph to manage the provenance graph as a collection of weakly connected sets. As discussed, this easily beats the recursive querying approach.

Few systems focus on capturing minimal volume of lineage data and optimizing the storage using domain properties and the detailed knowledge of transformations applied e.g., SubZero~\cite{Wu-icde13},  Anand et al.~\cite{anand-edbt09} etc. Our paper is domain-agnostic and is targeted towards the black-box lineage scenario wherein the lineage service does not have the details of internals of the transformations/UDFs being applied. Few systems e.g., ~\cite{Chapman,chen-icde08, Jagadish-tods90} start with the provenance data representation wherein the transitive closure of the provenance graph (i.e., for each data-item, its full provenance) is materialized and then propose techniques to reduce the storage cost. Our paper focuses on the scenario wherein the provenance data comprises of provenance triples capturing lineages across individual transformations. 
\vspace{-5pt}
\section{Conclusions}
\label{sec:conc}
\vspace{-3pt}

We proposed a provenance framework wherein we manage the workflow provenance graph as a collection of weakly connected sets, by exploiting the workflow dependency graph. The proposed approach is effective and provides significant speed-ups vis-a-vis existing recursive querying based methods.

\vspace{-5pt}
\small{
\bibliographystyle{IEEEtran}
\bibliography{provenance}
}

\end{document}